# Low peak-power pulse compression in gas-filled Herriott cells in the 2 µm wavelength range

Johann Gabriel Meyer[1,*], Felix Ritzkowsky[2], Fatemehsadat Ghaffari[2], Kevin Schwarz[1], Nazar Kovalenko[1], Christian Franke[3], Andrea Trabattoni[2,4,5], Oleg Pronin[1,3]

[1] *Helmut Schmidt University, Holstenhofweg 85, 22041 Hamburg, Germany*
[2] *Center for Free-Electron Laser Science CFEL, Deutsches Elektronen-Synchrotron DESY, Notkestraße 85, 22607 Hamburg, Germany*
[3] *n2-Photonics GmbH, Hans-Henny-Jahnn-Weg 53, 22085 Hamburg, Germany*
[4] *Leibniz University Hannover, Institute of Quantum Optics, Welfengarten 1, 30167 Hannover, Germany*
[5] *Cluster of Excellence PhoenixD (Photonics, Optics, and Engineering-Innovation Across Disciplines), 30167 Hannover, Germany*
*\* johann.meyer@hsu-hh.de*

**Abstract:** At laser wavelengths longer than the prominent 1 µm range of high-power ytterbium-doped lasers, nonlinear phase shifts produced in nonlinear media for spectral broadening and subsequent pulse compression decrease drastically. Consequently, at the 2 µm wavelength range, the threshold of the applicable peak power for pulse compression in gas-filled multipass cells increases. The common approach of choosing a Herriott multipass cell configuration close to the concentric resonator does not necessarily lead to the highest total nonlinear phase shift, due to a restriction of the total number of reflections on the cell mirrors of a given size. Therefore, an analytical approach is presented here to maximize the nonlinear phase shift for a given set of mirrors, considering lossless and dispersionless propagation. Furthermore, to achieve pulse compression with gas-filled multipass cells for relatively low peak powers at wavelengths around 2 µm, we developed a high-pressure gas cell and demonstrated experimentally pulse compression in the negative- and positive dispersion regimes, with achieved pulse durations of around 40 fs and 55 fs respectively.

## 1. Introduction

Spectral broadening in multipass cells has become an established technology in the 1 µm spectral range for the temporal compression of sub-ps long pulses from ytterbium-based laser systems. Post-compression methods enable the generation of ultrashort pulses with durations of a few 10 to sub-10 fs, that are similar to many Ti:sapphire based laser systems (wavelength around 0.8 µm) while providing higher repetition rates and average powers [1–6]. Novel IR laser technologies are attracting a large interest recently, showing the capability of extending the protocols of ultrafast lasers to longer wavelengths. In the 2 µm wavelength domain, Cr:ZnS laser oscillators offer the generation of ultrashort pulses with durations around 30 fs [7]. However, the average power scaling of Cr:ZnS comes with similar challenges as Ti:sapphire [8–10], and the amplification to higher energy pulses can lead to longer pulse durations [11]. Close to a wavelength of 2 µm, pulse energies in the µJ and mJ domains are provided by holmium and thulium based laser systems [12–20]. So far, multipass spectral broadening and compression for 2 µm laser pulses has been employed by only a few research groups. First, spectral broadening of 1.74 mJ pulses with a peak power of approx. 1.2 GW was demonstrated by Cankaya et al. [21]. Shortly after that, temporal pulse compression was successfully achieved by Gierschke et al. for 158 µJ pulses with a peak power of 1.2 GW from a thulium laser [22]. Eisenbach et al. demonstrated the temporal compression of 1.49 and 1.76 mJ pulses with peak powers of 11 and 15 GW from a thulium laser system [23]. In all three cases, gas-filled Herriott cells were used to compress 2 µm pulses with peak powers in the GW domain. However, laser systems operating at higher repetition rates and producing sub-100 µJ pulse energies, like some commercially available optical parametric amplifiers (OPA), deliver considerably lower peak powers. Suzuki et al. demonstrated excellent compression from 750 to 97 fs for 80 µJ pulses with a peak power of just 100 MW from their own Ho:CALGO regenerative amplifier, by employing a Herriott cell including a YAG plate as the nonlinear broadening medium [20]. The use of bulk nonlinear media is common in the 1 µm spectral range for sub-ps pulses with pulse energies in the range from sub-1 to 100 µJ [1,5,6,24]. For higher pulse energies, gas-filled multipass cells are often used [2,3,5,6]. However, even at lower pulse energies, gas-filled multipass cells are of interest, as bulk nonlinear media require reflective coatings, can easily be damaged, are sensitive to mode-matching, and often require dispersion management with customized dispersive cell mirrors. Kadiwala et al. realized the temporal compression of 15 µJ pulses at a wavelength of 1030 nm with a pulse duration of 270 fs and a peak power of 52 MW in a Herriott cell filled with 20 bar of krypton, while the beam pointing stability was already affected considerably [25]. As will be discussed in the following section, the spectral broadening in gas-filled Herriott cells becomes increasingly challenging at longer wavelengths. In this work, we theoretically and experimentally investigated the post-compression of femtosecond pulses with central wavelength around 2 µm and peak power in the 100 MW range by means of a compact gas-filled multipass cell. In this context, we investigate a parameter space of spectral broadening while lowering peak power and pulse energy. Therefore, we abandoned the common Herriott cell configurations close to the concentric resonator to allow for a higher nonlinear phase shift for the whole cell.

## 2. Design and theoretical consideration

When transitioning from 1 µm to 2 µm wavelengths, spectral broadening and temporal compression of ultrashort laser pulses in gases come with a few challenges. One limitation arises from the wavelength dependence of the nonlinear phase shift, which is responsible for spectral broadening. It will be discussed how to find a Herriott cell configuration to maximize the spectral broadening. Eventually, the spectral broadening can be increased further by increasing the pressure in the gas-filled Herriott cell. As will be discussed later, the smaller volume resulting from shorter cells could practically enable elevated pressures. Another limitation arises from the general wavelength scaling of the bandwidth of dispersive mirrors. The invisibility of the 2 µm radiation to the human eye and to silicon-based camera sensors poses an additional challenge to the alignment of the optical scheme.

### 2.1. Challenges for spectral broadening in gas-filled Herriott cells at the 2 µm wavelength range

First of all, the spectral broadening of ultrashort pulses is linked to the nonlinear phase shift produced via the nonlinear refractive index of the $\chi^{(3)}$-nonlinearity of the gas [26]. The nonlinear phase shift $\varphi_{nl}$ itself scales explicitly with the reciprocal of the wavelength $\lambda$:

$$\varphi_{nl} = \frac{2\pi}{\lambda} \int n_2 \cdot I(z)\, dz \tag{1}$$

Here, $n_2$ denotes the nonlinear refractive index. Furthermore, the nonlinear phase shift $\varphi_{nl}$ is linearly dependent on the intensity of the laser pulses $I(z) \propto P/A$, which scales inversely with the beam area $A$. The beam area is affected by the configuration of the Herriott cell. Inside the Herriott cell, the laser beam is reflected repetitively by two cell mirrors that form a stable resonator. Usually, the propagating beam is matched to the fundamental Gaussian eigenmode of this resonator. For a given resonator configuration, the mode area is scaling, at all positions $z$, linearly with the wavelength ($A \propto \lambda$). Additionally, as the beam radius inside the Herriott cell scales with the square root of the wavelength ($w \propto \sqrt{A} \propto \sqrt{\lambda}$), the number of beams that fit on the cell mirrors of the Herriott cell scales inversely with the square root of the wavelength (later discussed, see Eq. 5 and Fig. 2). Overall, it can be expected that the nonlinear phase shift in a Herriott cell scales with the wavelength as $\varphi_{nl} \propto \lambda^{-2.5}$ [27]. For the same Herriott cell's resonator geometry, it would follow that laser pulses with a wavelength of 2 μm would need a roughly 6 times higher peak power to achieve the same nonlinear phase shift and spectral broadening when compared to laser pulses with a wavelength of 1 μm. To achieve sufficient spectral broadening in gas-filled Herriott cells with 2 μm laser pulses of similar pulse energies as typically employed in the 1 μm spectral range, it becomes crucial to optimize the Herriott cell for a maximal total nonlinear phase shift. An analytical approach to solving this problem, under the assumption of negligible second-order dispersion, is presented in the following section. It turns out that it is not sufficient to simply choose the longest resonator configuration with the highest nonlinear phase shift for each individual pass between the cell mirrors. The resulting configuration and nonlinear phase shift are presented in equations 2, 7, 8, and 9.

An additional complication can arise from a decreased bandwidth of the cell mirrors at longer center wavelengths. However, independent of the center wavelength, the frequency bandwidth required from the mirrors is directly proportional to the inverse of the pulse duration after spectral broadening and pulse compression. This could limit the pulse compression to longer pulses at a wavelength of 2 μm.

The mirrors of a Herriott cell are often dielectric mirrors to get a high reflectivity and a high damage threshold. However, the supported spectral bandwidth of a dielectric Bragg mirror made from a stack of two dielectric materials with a certain refractive index contrast is proportional to the center frequency. For longer wavelengths, the center frequency decreases, and the covered spectral bandwidth decreases correspondingly. Unless materials combinations like $Si/SiO_2$ with a high refractive index contrast, but a reduced damage threshold, are employed, the supported Fourier-transform limited (FTL) pulse durations of such Herriott cells could be about a factor two longer at a wavelength of 2 μm when compared to common Herriott cells for spectral broadening at a wavelength of 1 μm.

*2.2. An analytical approach to find the configuration with the highest nonlinear phase shift*

Here, we suggest an analytical approach to find the configuration for a gas-filled Herriott cell that produces the highest nonlinear phase shift and the strongest spectral broadening. Within this approach, it is assumed that the dispersion of the gas has a negligible impact on the propagating laser pulses. Furthermore, it is assumed that the Herriott cell mirrors introduce no relevant dispersion. However, for long propagation lengths, high gas pressures, and many mirror reflections, these assumptions can become invalid. Pulses stretched by dispersion could lead to significantly reduced nonlinear phase shifts and spectral broadening. Dispersion compensation of the gas can become tricky, as it would require precise control of the mirror dispersion. This could limit the spectral broadening to certain gas pressures. Therefore, the considerations presented here might be seen as a theoretical limit. Still, this limit could be exceeded if spectral broadening in the negative dispersion regime were considered. In this case, the output might depend strongly on the input pulse parameters and the dispersion. Furthermore, nonlinear pulse break-up and damaged mirrors could be the result of strong nonlinear effects during the self-compression in the negative dispersion regime.

Typically, the nonlinear phase shift from a single pass between the two mirrors of a symmetric Herriott cell increases with the distance between the cell mirrors, until the cell configuration approaches the concentric resonator. Tight focusing of the resonator eigenmode allows for the highest intensities and therefore for the strongest nonlinear phase shift. The on-axis nonlinear phase shift for $2N$ passes during the propagation in a Herriott cell with $N$ reflections on each mirror, can be calculated in good approximation analytically if pulse stretching due to dispersion can be neglected [6,28]:

$$\varphi_{nl,tot} = 2N \cdot \varphi_{nl,pass} = 2N \cdot \frac{\pi^2}{2} \frac{P_{peak}}{P_{crit}} \frac{M}{N} = \pi^2 \frac{P_{peak}}{P_{crit}} \cdot M \tag{2}$$

With the peak power of the laser pulses $P_{peak}$, a critical peak power defined approximately as $P_{crit} := \lambda^2/(8nn_2)$, the refractive index $n$, and the nonlinear refractive index $n_2$. $N$ is the number of beam reflections on each cell mirror, i.e., the number of roundtrips in the Herriott cell. The configuration parameter $M$ $(1 \ldots N-1)$ defines how many spots the spot pattern on each mirror progresses for a roundtrip [Fig. 1]. The progression angle between consecutive reflections is simply given by $\vartheta = 2\pi \cdot M/N$.

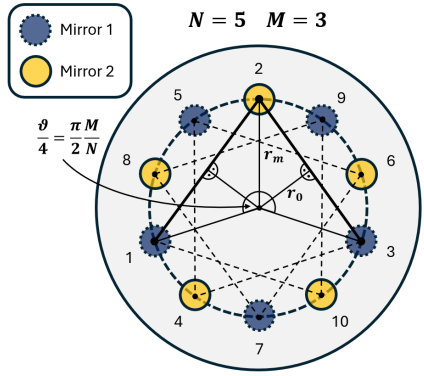

Fig. 1. (cf. [27]) Visualization of the spot pattern formation on the mirrors of a symmetric Herriott cell. The viewing direction of this projection is along the cell axis. The spots of beam reflections on mirror 1 are indicated in blue, and the spots on mirror 2 are yellow. In this example, there are five spots on each mirror ($N = 5$) and the next spot on the same mirror appears as the third neighbor ($M = 3$). All spots are numbered in the order in which the reflections appear during the beam propagation in the Herriott cell. The distance between the central axis of the cell and the spots on the mirrors is marked as the pattern radius on the mirrors $r_m$. The closest distance between the propagating beams and the cell axis is marked as the pattern radius at the center of the cell $r_0$.

The parameters $N$ and $M$ are linked to the radius of curvature of the cell mirrors $R$ and the distance between the cell mirrors $L$ [29]:

$$L = 2\sqrt{R^2 - r_m^2} \cdot \sin^2\left(\frac{\pi M}{2 N}\right) \approx 2R \cdot \sin^2\left(\frac{\pi M}{2 N}\right) \tag{3}$$

With radius of the spot pattern on the cell mirrors $r_m$ [Fig. 1]. It becomes obvious from Eq. 2 that the nonlinear phase shift for one single pass becomes maximal for $M = N - 1$, where the cell length approaches $L \to 2R$ (concentric resonator) for large $N$ [Eq. 3]. For a given number of reflections $N$, also the total nonlinear phase shift in the gas-filled Herriott cell is maximized for $M = N - 1$ [Eq. 2] [28]. However, in the limit of the concentric resonator, the beam radius on the cell mirrors diverges. This radius can be approximated as [27,28]:

$$w_m \approx \sqrt{\frac{\lambda}{\pi} R \cdot \tan\left(\frac{\pi M}{2 N}\right)} \tag{4}$$

Eventually, the beam radius on the cell mirrors $w_m$ decides how many spots can be arranged on a mirror of a given size [Fig. 2]. The spots on the mirror are located on a circle with a pattern radius $r_m$. This radius is usually limited by the size of the cell mirror. The maximum number of spots $N$ is reached when the beams surrounded by a safety margin, which is $x_s$ times larger in diameter than the beam diameter, fill out the circle of the spot pattern [27]:

$$2\pi \cdot r_m \approx N \cdot x_s \cdot 2w_m \tag{5}$$

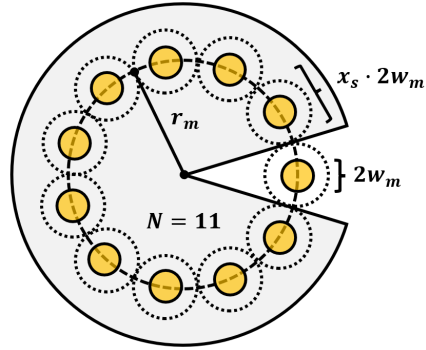

Fig. 2. [27] Visualization of the no-clipping condition for the spot pattern on one Herriott cell mirror [Eq. 5]. Each beam, with a radius on the mirror $w_m$, is surrounded by a $x_s$ times larger circle to avoid an overlapping of the beams and clipping during the coupling into and out of the Herriott cell. Given a certain tolerable alignment safety factor $x_s$, the number of spots $N$ is limited by the circumference of the circle on which the spots are lying. Note that a circle that is 1.5 times larger than the 1/e² radius of a Gaussian beam contains just 99 % of the optical power. Typically, $x_s$ would be between 2 and 3 to avoid clipping during small misalignments.

To maximize the total nonlinear phase shift [Eq. 2], we should not limit ourselves to maximum nonlinear phase shift for a single pass, which appears for $M = N - 1$. We should allow for shorter resonator configurations to reduce the beam radius on the mirrors, and to enable more passes, which could result then in a higher total nonlinear phase shift. For the maximization of Eq. 2, we need to find the largest possible parameter $M$, while we keep $N$ as a free parameter. The equation for the beam radius on the mirrors [Eq. 4] and the no-clipping equation [Eq. 5] can be solved for the parameter $M$ [27]:

$$M = N \cdot \frac{2}{\pi} \cdot \arctan\left(\frac{\pi^3 \, r_m^2}{\lambda \, R \, x_s^2 \, N^2}\right) \tag{6}$$

In Fig. 3, the possible values for the configuration parameter $M$ are plotted against the number of reflections on each cell mirror $N$, respectively the number of roundtrips, according to Eq. 6. Note that $M$ is monotonously increasing with the pattern radius $r_m$, but monotonously decreasing with the wavelength $\lambda$, with the radius of curvature of the mirrors $R$, and with the alignment safety factor $x_s$ [Eq. 5]. With respect to the number of reflections $N$, the configuration parameter $M$ is first increasing. After reaching a global maximum, $M$ is monotonously decreasing with $N$.

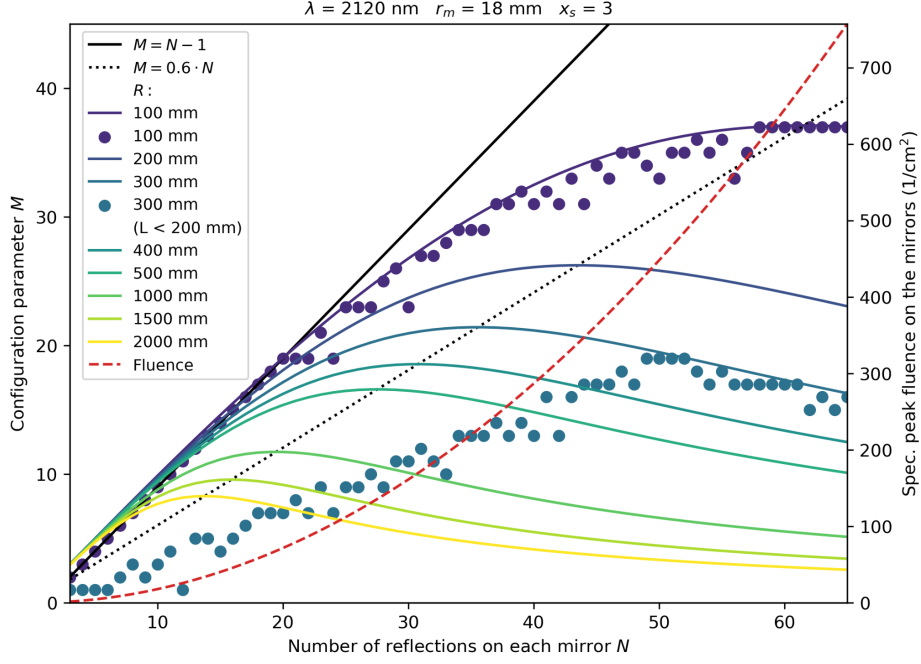

Fig. 3. The dependency between the number of beam reflections on each Herriott cell mirror $N$ and the cell configuration parameter $M$ is plotted according to Eq. 6 for different radii of curvature of the Herriott cell mirrors $R$, for a given wavelength of $\lambda = 2150$ nm, for a pattern radius $r_m = 18$ mm on mirrors with a diameter of 50 mm, and for a conservative alignment safety factor of $x_s = 3$. For the radii of curvature $R$ of 100 mm and 300 mm, the highest integer values of $M$, limited by Eq. 6, was evaluated, while solutions with common prime factors for $M$ and $N$ were eliminated. Furthermore, for $R = 300$ mm, it was considered that the Herriott cell should not exceed a length of 200 mm [Eq. 3] to fit into the here designed high pressure chamber (max. 30 bar). Again, for unlimited lengths and for smaller $N$, the longest configuration with $M = N - 1$, closest to the concentric resonator $(L \to 2R)$ [Eq. 3], is typically accessible. However, for larger $N$, clipping on the cell mirrors due to an increasing beam radius prohibits this configuration [Eq. 5]. Still $M$ is increasing with $N$, until a global maximum is reached, specified by equations 7 and 8. For each radius of curvature, the corresponding maximum is lying on the dotted line with $M/N \approx 0.6$ [Eq. 9]. Furthermore, with increasing $N$, the fluence on the Herriott cell mirrors is increasing quadratically according to Eq. 11, which is given here as a specific Fluence, divided by the pulse energy.

After a suitable variable substitution, numerical analysis of Eq. 6 reveals the location of the maximum of the configuration parameter $M$ to be at [27]:

$$N_{max-M} = \sqrt{\frac{\pi^3 \, r_m^2}{\lambda \, R \, x_s^2}} \cdot 0.511 \cdots \tag{7}$$

At this global maximum, the configuration parameter assumes a value of [27]:

$$M_{max} = \sqrt{\frac{\pi^3 \, r_m^2}{\lambda \, R \, x_s^2}} \cdot 0.848 \cdots \tag{8}$$

At this maximum, the ratio of the parameters $M$ and $N$ is calculated with equations 7 and 8 to be [27]:

$$\frac{M_{max}}{N_{max-M}} = 0.603 \cdots \tag{9}$$

This fixed ratio of about 0.6 characterizes the resonator configuration by Eq. 3 for all Herriott cells with the largest configuration parameter $M$, respectively with the largest nonlinear phase shift in a dispersionless gas-filled Herriott cell. This configuration corresponds to a ratio of $L/R \approx 1.32$ [Eq. 3 and 9]. When a practical configuration is chosen it needs to be considered that the parameters $M$ and $N$ are positive integers and they should not have a common prime factor to get a complete pattern formation in the Herriott cell [28].

The proposed configuration of the maximum parameter $M$ leads typically to more passes on the cell mirrors than common configurations, which are closer to the concentric resonator [Fig. 3, Eq. 7]. Therefore, the higher number of reflections on the cell mirrors can reduce the total transmission of the cell. Considering a certain mirror reflectivity $\rho$, and summing over the nonlinear phase shift for each pass [Eq. 2], the total on-axis nonlinear phase shift can be evaluated with the geometrical series ($\rho < 1$):

$$\varphi_{nl,tot} = \sum_{n=0}^{2N-1} \varphi_{nl,n} = \sum_{n=0}^{2N-1} \frac{\pi^2}{2} \frac{P_{peak}}{P_{crit}} \frac{M}{N} \cdot \rho^n = \frac{\pi^2}{2} \frac{P_{peak}}{P_{crit}} \frac{M}{N} \cdot \frac{1-\rho^{2N}}{1-\rho} \tag{10}$$

Another trade-off arises from an increasing fluence on the cell mirrors when decreasing the beam radius on the mirrors to maximize the nonlinear phase shift. For large enough pulse energies, this can bring the fluence beyond the damage threshold of the mirrors. However, the presented approach of maximizing the nonlinear phase may not be necessary for such high pulse energies after all.

From the no-clipping condition for the beam radius on the mirrors [Eq. 5], the peak fluence of a Gaussian beam on the mirrors can be calculated [27]:

$$F_{peak,m} = \frac{2 E_p}{\pi w_m^2} \approx E_p \cdot \frac{2 x_s^2}{\pi^3 r_m^2} \cdot N^2 \tag{11}$$

It becomes obvious that the fluence scales quadratically with the number of beam spots on each mirror $N$. The specific fluence, divided by the pulse energy is included in Fig. 3. Depending on the actual pulse energy, a compromise between maximizing $M$, respectively the total nonlinear phase shift, and limiting the fluence on the mirrors might need to be found. From Fig. 3 and Eq. 2, it can be seen that with a radius of curvature of 100 mm, the configuration of max. nonlinear phase shift (max. $M$) could practically lead to an about two times higher nonlinear phase shift than the longest cell configuration ($M = N - 1$), while the fluence increases about a factor six. As the slope for $M$ decreases towards the maximum, a slightly lower $M$ than the maximum value can be chosen at a smaller $N$, while the fluence decreases already drastically.

Another advantage of choosing $M$ values that correspond to shorter cells (cf. Eq. 3) than the longest configuration ($M = N - 1$) is that it could allow to increase the gas pressure in the Herriott cell to higher values, which could also lead to higher total nonlinear phase shifts and therefore to stronger spectral broadening. Typically, the Herriott cell is placed into a pressurized gas vessel, which is filled with a gas with a certain nonlinear refractive index. The nonlinear refractive index is effectively increased by increasing the gas pressure, which could be used to increase the nonlinear phase shift to the desired amount. However, for safety reasons, the product of gas pressure and gas volume would be practically limited. Consequently, high gas pressures for sufficient spectral broadening have to be afforded by small volumes of the pressure vessel surrounding the Herriott cell. Shorter Herriott cells can be beneficial here to reduce the volume and maximize the pressure. Compared to the longest configuration with

$M = N - 1$, and $M/N \approx 1$, close to the concentric resonator ($L = 2R$), it can be seen from Eq. 3 that the configuration of maximal $M$ (max-$M$) has with $M/N \approx 0.6$ [Eq. 9] a roughly 1.5 times shorter cell length, which would allow for a 1.5 times higher gas pressure for a given pressure-volume-product. This could result in an additional increase of the nonlinear phase shift by a factor of 1.5, unless the second order dispersion starts to stretch the laser pulses significantly over the long effective propagation length in the high gas pressure. Combined with the already about two times higher nonlinear phase shift expected for the max-$M$ configuration at $R = 100$ mm in Fig. 3, it could result in a theoretical increase of the nonlinear phase shift by a factor of three compared to the longest configuration. Conversely, this might reduce the necessary peak power for a given spectral broadening factor by up to a factor three [26].

So far, the Herriott cell configuration was specified by the parameters $M$ and $N$ [Eq. 7 and 8], respectively by the corresponding resonator configuration, given by the ratio $L/R$ [Eq. 3 and 9]. In addition, it is relevant to define a relation between the dimension of the cell mirrors and the total nonlinear phase shift. The diameter of the cell mirrors has, via the achievable pattern radius $r_m$, a direct impact on the number of passes that the pattern can hold within the no-clipping condition [Eq. 5 and 6]. If we limit ourselves already to the previously introduced configuration of max. $M$, this condition is reflected in the linear scaling of $M_{max}$ with respect to the pattern radius $r_m$ [Eq. 8]. On the other hand, $M_{max}$ is scaling inversely with the square root of the radius of curvature of the cell mirrors $R$. Decreasing $R$ has practical limitations for a given mirror diameter. If we consider a fixed minimal ratio of $r_m/R$, we will end up with a scaling of $M_{max} \propto \sqrt{r_m/R \cdot r_m}$ [Eq. 8]. However, for a fixed ratio of $r_m/R$, the volume of a simple gas vessel surrounding the Herriott cell would roughly scale with $r_m^3$, if the scaling of the opto-mechanics is not limiting. This would lead to a scaling of the max. gas pressure with $r_m^{-3}$, for a fixed pressure-volume product $p \cdot V$. As the nonlinear refractive index should scale linearly with the gas pressure $p$, it would result in a scaling of the total nonlinear phase shift according to [Eq. 2 and 8]:

$$\varphi_{nl,tot} \propto p \cdot M_{max} \propto \frac{p \cdot V}{r_m^3} \cdot \sqrt{\frac{r_m}{R} \cdot r_m} \propto \frac{1}{r_m^{2.5}} \qquad (12)$$

While this scaling suggests the use of cell mirrors with small diameters, going in hand with a small pattern radius $r_m$, it needs to be considered that this scaling is limited by mechanical design constraints, by practically usable gas pressures, the tolerable nonlinear phase shift per pass, and the fluence on the mirrors [Eq. 7 and 11]. The peak fluence on the mirrors would scale in the max-$M$ configuration, with $M/N \approx 0.6$ and with a fixed ratio $r_m/R$, as $F_{peak,m} \propto 1/w_m^2 \propto 1/R \propto 1/r_m$ [Eq. 4, 9, and 11].

On the other hand, we could consider a given limit for the pressure in the gas vessel. Furthermore, we start from a given Herriott cell with a certain mirror diameter, for which we already minimized the radius of curvature $R$, or maximized the ratio $r_m/R$ to its practical limit. To increase the nonlinear phase shift according to Eq. 2 and 8, it would only make sense to further increase the cell mirror diameter, together with the pattern radius $r_m$, until a certain limiting pressure-volume product is reached for the correspondingly increased cell volume.

If the total propagation distance in the Herriott cell and the gas pressure are high enough, gas dispersion will stretch the pulses. To minimize the effect of the gas dispersion, it could be helpful to look at the scaling of the group delay dispersion (GDD) of the gas for a given ratio of $r_m/R$. With equations 3, 9, and 12, it follows a scaling of the gas dispersion for the max-$M$ configuration:

$$GDD_{gas} \propto p \cdot 2N_{max-M} \cdot L \approx 2 \cdot \frac{p \cdot M_{max}}{0.6} \cdot 2R \cdot \sin^2\left(\frac{\pi}{2} \cdot 0.6\right) \propto \frac{1}{r_m^{2.5}} \cdot \frac{R}{r_m} \cdot r_m \propto \frac{1}{r_m^{1.5}} \qquad (13)$$

Once the gas dispersion stretches the pulses, and dispersion compensation is impractical, this scaling would suggest considering larger mirror diameters to reduce the GDD of the gas. If all these scaling considerations did not result in sufficient spectral broadening for a given pulse energy and given input pulse duration, or led to an excessive amount of complexity, it would still be possible to consider a solid broadening medium instead of the gas broadening approach discussed here [1,20,24].

## 3. Experimental setup and results

### 3.1. Laser source

For the current experiment, IR laser pulses were generated in a commercial optical parametric amplifier (OPA) (*Orpheus, Light Conversion*). The OPA provided pulses with tunable wavelength around 2 µm and pulse energy approximately between 50 and 70 µJ, upon being pumped at 1035 nm with 160 fs pulses at a repetition rate of 10 kHz (*Pharos, Light Conversion*). The OPA was operated at two distinct central wavelengths, namely 2000 and 2120 nm, thus slightly below or above the degeneracy point of the OPA at 2070 nm, respectively. The two operating OPA wavelengths coincided with two different group delay dispersion (GDD) regions of the mirrors. In particular, the mirrors induced an average positive GDD to the 2000 nm pulses while imprinting a negative GDD to the 2120 nm ones. The post-compression was studied in both regimes, leading to peculiar features.

In particular, for the 2120 nm OPA pulses, we could observe self-compression after the Herriott cell. For the 2000 nm pulses, instead the Herriott cell was operated in a normal dispersion regime. For an emission wavelength of 2120 nm, a spectral bandwidth of 33 nm (FWHM) and a pulse duration of 197 fs was measured [Fig. 4]. However, the spectrum itself would support a Fourier-transform limited pulse duration of about 160 fs. The spectral chirp of the pulses was estimated to be roughly +6700 fs$^2$.

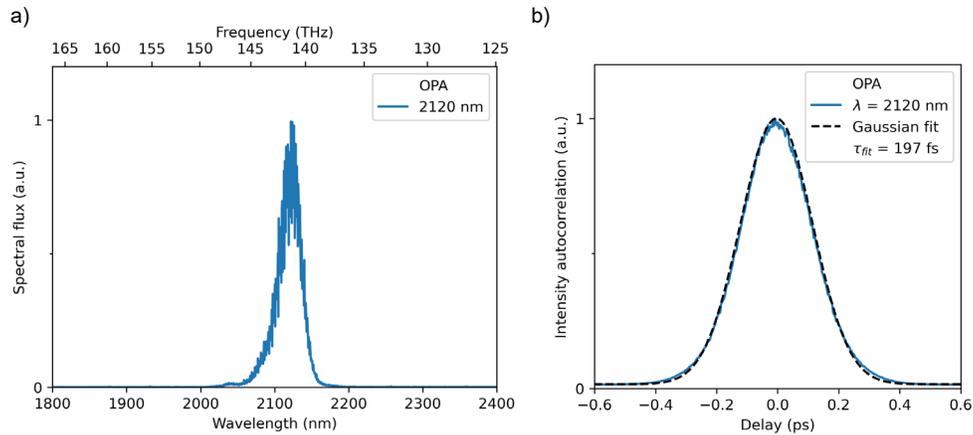

Fig. 4. a) Measured spectrum of the output beam of the OPA with a central wavelength of 2120 nm. The Fourier-transform-limited (FTL) pulse duration supported by this spectrum corresponds to about 160 fs. b) The corresponding pulse duration was estimated to be 197 fs with an intensity autocorrelation measurement. The increased pulse duration and the Gaussian shape of the autocorrelation suggest a quadratic phase in the spectrum.

*3.2. Setup of the Herriott cell for spectral broadening and pulse compression*

As described in the previous theory section, the configuration of the Herriott cell was designed to achieve a large nonlinear phase shift in gaseous nonlinear media and thus enable a sufficient spectral broadening for rather low pulse energies and peak powers.

Assuming pulse energies around 50 µJ and input pulse durations around 200 fs, we considered cell mirrors with a diameter of 50 mm (~2 inch) to get a large configuration parameter $M$ according to Eq. 8, and correspondingly a strong spectral broadening from a high nonlinear phase shift, according to Eq. 2. For this mirror diameter, radii of curvature down to $R = 100$ mm are practically available. The possible cell configurations for this radius of curvature and higher values are depicted in Fig. 3. The mirror diameter would allow a pattern radius of at least $r_m = 18$ mm [Fig. 2], without clipping on the mirror edge. An alignment safety factor of $x_s = 3$ was considered to avoid clipping of the 2 µm radiation [Fig. 2]. The highest value of $M = 36$ is reached for a number of reflections of $N = 61$ [Eq. 7 and 8]. However, as the losses on the mirrors and the fluence increase with each additional reflection, a slightly smaller $M$ might be chosen, while reducing $N$ already significantly [Fig. 3]. For $N = 51$ and a pulse energy of 50 µJ, the fluence would reach an acceptable value of 23 mJ/cm².

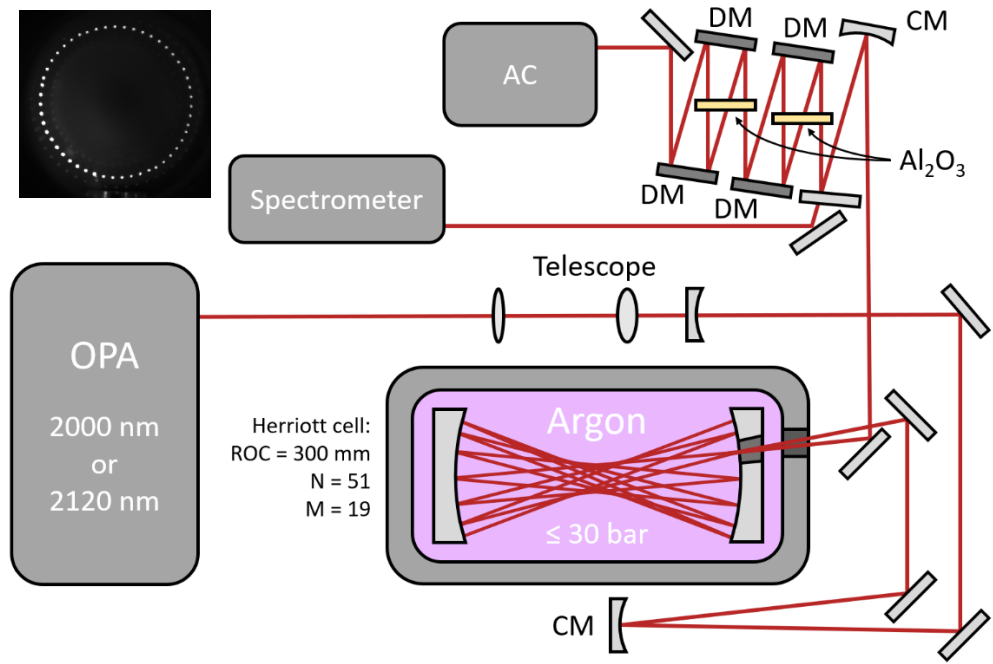

Fig. 5. Schematic setup for spectral broadening laser pulses at wavelengths around 2 µm in a gas-filled Herriott cell. The OPA was emitting a pulsed laser beam at a wavelength of 2000 or 2120 nm and a repetition rate of 10 kHz. With a telescope, the beam was mode-matched (for 2120 nm) to the Herriott cell. Then, the beam was focused by a curved mirror (CM) with a ROC of 1 m. The Herriott cell was aligned in a configuration with the parameters $M = 19$, $N = 51$ (see inset), and a mirror distance of $L = 183$ mm. The Herriott cell was placed into a gas vessel that allowed us to fill argon with a pressure of up to 30 bar. A 5 mm thick AR-coated sapphire plate was used as a laser window. After the cell, the beam was collimated by another curved mirror (ROC = 1 m). For the temporal compression of the beam, dispersive mirrors (DM) and sapphire plates ($Al_2O_3$) could be added before measuring the pulse duration with an intensity autocorrelator (AC).

Based on this mirror choice, a gas vessel was designed that would fit the Herriott cell, even with the maximal mirror separation of about 200 mm for the longest configuration [Eq. 3]. The designed vessel allows technically for a gas pressure of about 30 bar. Already with the inexpensive nonlinear medium argon, considerable spectral broadening would be expected. Principally, it would be possible to double the nonlinear refractive index with krypton, but the group delay dispersion, leading eventually to temporal stretching, would also double [28]. For argon, the group delay dispersion of about $+8 \cdot 10^{-3}$ fs$^2$/mm [30] would accumulate over a total propagation length of about 16.4 m (for $R = 100$ mm, $r_m = 18$ mm, $M = 36$, $N = 61$, Eq. 3) and with a gas pressure of 30 bar to a GDD of about +3900 fs$^2$, which could stretch a 200 fs flat-phase Gaussian pulse only slightly to about 207 fs [31].

However, due to availability issues, we had to rely on mirrors with a radius of curvature of $R = 300$ mm. As the manufactured pressure vessel allowed only a mirror separation of about $L = 200$ mm, it was not possible to fit the configuration of max-$M$ ($M/N \approx 0.6$) [Eq. 9], which would have required a separation of 392 mm [Eq. 3]. Furthermore, to keep a constant pressure-volume product, the larger pressure vessel would have only allowed for a pressure of 15 bar, reducing the achievable nonlinear phase shift. Therefore, we continued with the produced pressure vessel, and chose a shorter configuration with $M = 19$, $N = 51$ from Fig. 3, resulting in a mirror separation of $L = 183$ mm [Eq. 3]. The Herriott cell was aligned and mode-matched with a telescope for a wavelength of 2120 nm [Fig. 5]. The collimated beam was focused into the Herriott cell with a spherical mirror with a radius of curvature (ROC) of 1 m.

However, when the OPA was later tuned to a wavelength of 2000 nm, issues with the alignment procedure did not allow it to adjust and verify the alignment of the Herriott cell mirrors to the new mode. This could have led to the later observed problems with the beam profile at this wavelength [Fig. 6].

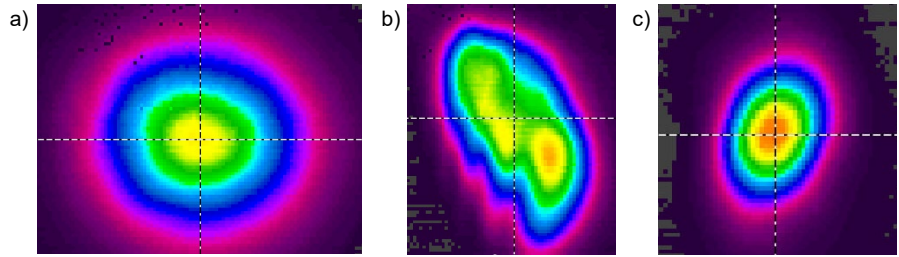

Fig. 6. a) Measured beam profile of the OPA at a wavelength of 2000 nm. b) Measured beam profile after the Herriott cell was filled with 30 bar of argon at a wavelength of 2000 nm and an input pulse energy of 45 µJ. c) Measured beam profile after the Herriott cell was filled with 30 bar of argon, and after the wavelength was changed to 2120 nm.

Overall, the lenses and mirror between the OPA and the Herriott cell introduced losses of about 20 %, which are attributed to imperfect lens and mirror coatings. The Herriott cell itself introduced losses of about 15 %, which would correspond to 101 reflections on cell mirrors with a reflectivity of 0.9984 %. The cell transmission was not affected significantly after filling the cell with 30 bar of argon.

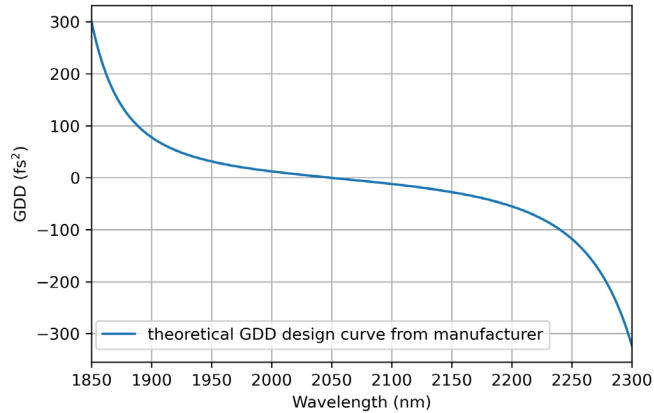

Fig. 7. This curve shows for each wavelength the theoretical group delay dispersion (GDD) introduced by one bounce on the employed Herriott cell mirrors ($R = 300$ mm), as it was specified by the design curve of the manufacturer. While the GDD is supposed to be 0 $fs^2$ for the design wavelength of 2050 nm, it deviates significantly at the other wavelengths, due to third-order dispersion (+530 $fs^3$ at 2050 nm), and dispersion of higher orders. At the operational wavelengths of 2000 nm and 2120 nm, the GDD was calculated to be +12 $fs^2$ and -18 $fs^2$ respectively.

The employed Herriott cell mirrors were originally designed for a center wavelength of 2050 nm with vanishing group delay dispersion at this wavelength [Fig. 7]. At the OPA wavelengths of 2000 and 2120 nm, the group delay dispersion deviates already noticeably from 0 $fs^2$. While a single bounce on these mirrors would not affect the pulse duration, the effect accumulates over the large required number of passes for the considered configuration. At an OPA wavelength of 2120 nm, the cell mirrors are expected to introduce a dispersion of -18 $fs^2$ per reflection, accumulating to -1818 $fs^2$ over 101 reflections. Depending on the actual dispersion of the real mirrors, the gas pressure, and the initial chirp of the laser pulses, the pulse propagation and spectral broadening could enter a regime of zero dispersion or even negative dispersion. Self-compression and strong nonlinear effects would be the result. However, small absolute deviations from the mirror's design curve could lead to relatively large deviations in the accumulated value, which makes it difficult to control this regime at a given gas pressure.

On the other side, at a wavelength of 2000 nm, the mirrors are expected to introduce a small positive dispersion of +12 $fs^2$ per reflection, which would add to +1212 $fs^2$ for 101 reflections. For the propagation of the laser pulses inside the Herriott cell, the positive mirror dispersion will add to the positive dispersion of the gas medium itself. This should lead to spectral broadening in the positive dispersion regime.

After the Herriott cell, the outgoing beam was recollimated with another curved mirror (ROC = 1 m). Then, the beam was either sent into a spectrometer (*APE waveScan*) or into an intensity autocorrelator (*APE pulseCheck*). To temporally compress the spectrally broadened pulses, a variable amount of dispersive mirrors and AR-coated sapphire plates were available.

### 3.3. Spectral broadening and self-compression at a wavelength of 2120 nm

When the wavelength of the OPA was tuned to a wavelength of 2120 nm, it was observed that the output pulses after the Herriott cell were often already temporally compressed without the addition of dispersive elements. Over the propagation within the cell, the initial chirp, the dispersion of the gas, and the self-phase modulation (SPM) in the Herriott cell must have been compensated by the cell mirrors, which led to the appearance of self-compression. At an argon pressure of 8 bar inside the Herriott cell, pulses with an energy of about 50 µJ were sent into the Herriott cell. They experienced strong spectral broadening, resulting in a spectrum that supports 40 fs (FTL) short pulses [Fig. 8]. Directly after the Herriott cell, the pulse duration was evaluated with an autocorrelator. A Gaussian fit of the center region of the autocorrelation indicates a pulse duration of 46 fs [Fig. 8]. However, the appearance of side pulses does also increase the width of the center region. Therefore, this duration can be considered as an upper limit for the main pulse. Practically, additional positive or negative dispersion did not decrease the autocorrelation width. The center peak of the autocorrelation was surrounded by a pedestal, which could have been caused by a side peak. Such side peaks might appear due to strong nonlinear effects in the self-compression regime. For higher pressures, it was observed that the number and intensity of the side peaks increased in the autocorrelation. This excludes the use of higher gas pressures in this broadening regime. The ideal self-compression at each pressure would depend on a delicate balance of dispersion, nonlinearity, and pulse energy within the Herriott cell.

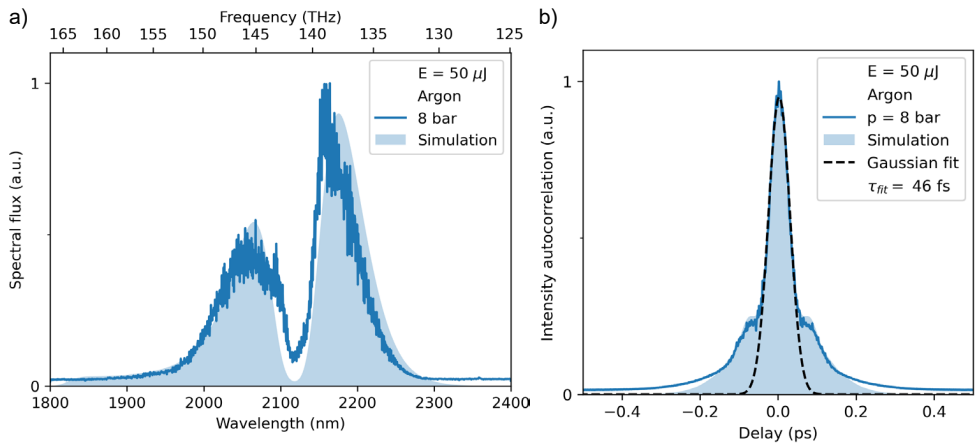

Fig. 8. a) Measured spectrum of the output beam of the Herriott cell for an OPA center wavelength of 2120 nm, and a pulse energy of about 50 µJ before the Herriott cell. The Herriott cell contained an argon pressure of 8 bar. The Fourier-transform limited (FTL) pulse duration supported by this spectrum corresponds to about 40 fs. The spectrum resulting from the numerical simulation of the spectral broadening in the Herriott cell is added in the background. b) The pulse duration was evaluated with an intensity autocorrelator without adding any dispersive elements after the cell. A Gaussian fit of the central part of the autocorrelation assumed a pulse duration of 46 fs (FWHM). For the autocorrelation trace calculated from the simulated pulse shape, the pulse duration assumed a value of 36 fs (FWHM).

To support the experimental observation, we conducted a numerical simulation of the self-compression in the Herriott cell. Therefore, we modelled the Herriott cell as a continuous fiber (*FiberDesk*). A Gaussian input pulse with a pulse duration of 197 fs and an initial chirp of +6700 fs$^2$ was assumed for the simulation. Furthermore, the mirror dispersion up to the 12$^{th}$ order was extracted at a wavelength of 2120 nm by a polynomial fit from the theoretical design curve [Fig. 7]. To fit the simulation to the experimentally observed autocorrelation, the second-order dispersion of the Herriott cell mirrors was adjusted manually to a value of roughly -42 fs$^2$ per reflection, which deviates from the theoretical value of -18 fs$^2$ [Fig. 7]. While the mirror's negative second-order dispersion overcompensates the +12 fs$^2$ per pass from 8 bar of argon, it does not compensate completely for the positive chirp of the OPA pulses. Still, the simulation resulted in a FWHM pulse duration of 36 fs. The pulse shape shows a clear breakup of the self-compressed laser pulse, which explains the pedestal in the measured autocorrelation [Fig. 8].

When increasing the pressure in the simulation, it was observed that the second pulse becomes stronger.

*Spectral broadening and pulse compression in the positive dispersion regime at a wavelength of 2000 nm*

When the emission wavelength of the OPA is tuned to 2000 nm, spectral broadening in the positive dispersion regime will appear in the Herriott cell. Already for an argon pressure of 10 bar in the Herriott cell, strong spectral broadening was observed for a pulse energy of 58 µJ, resulting in a spectrum with a theoretical Fourier-transform limited pulse duration of 56 fs [Fig. 9]. The uncompressed pulse duration was evaluated to be 236 fs from the Gaussian fit of the measured autocorrelation trace [Fig. 9]. The good agreement of the Gaussian fit and the increase of the pulse duration directly after the cell is typical for spectral broadening in the positive dispersion regime. Good compression results are expected after the application of negative group delay dispersion.

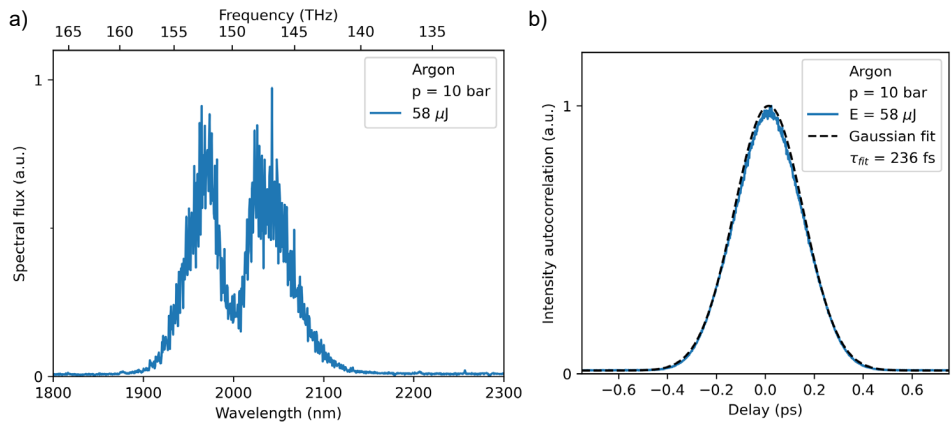

Fig. 9. a) Measured spectrum of the output beam of the Herriott cell for an OPA center wavelength of 2000 nm, and a pulse energy of 58 µJ before the Herriott cell. The Herriott cell contained an argon pressure of 10 bar. The Fourier-transform limited pulse duration supported by this spectrum corresponds to about 56 fs. b) The pulse duration was evaluated with an intensity autocorrelator without adding any dispersive elements after the cell. A Gaussian fit of the autocorrelation assumed an uncompressed pulse duration of 236 fs (FWHM).

Before attempting the compression of the spectrally broadened laser pulses, the argon pressure was increased to 30 bar to maximize the nonlinear phase shift and to investigate how much the pulse energy can be reduced. Therefore, the spectrally broadened spectra were recorded for different energies ranging from 7 to 69 µJ [Fig. 10].

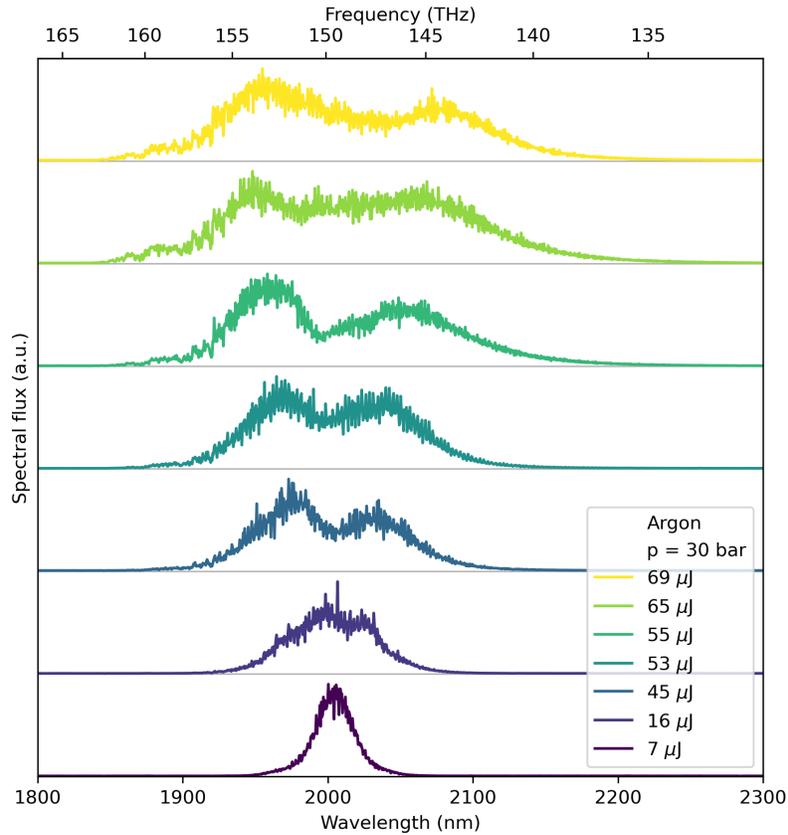

Fig. 10. At an OPA center wavelength of 2000 nm and an argon pressure of 30 bar, the broadened output spectra were measured after the Herriott cell for different input pulse energies.

For the highest pulse energy of 69 µJ, the broadened spectrum supports a Fourier-transform limited pulse duration of 43 fs [Fig. 11]. The temporal compression of the laser pulses was realized by a variable number of reflections on dispersive mirrors which introduce about -300 fs$^2$ per reflection, and by the propagation through several 5 mm thick plates of sapphire, which exhibit a negative group delay dispersion of about -120 fs$^2$/mm. After eight mirror reflections and 30 mm of sapphire, the pulse duration was compressed to 55 fs according to the Gaussian fit of the measured autocorrelation [Fig. 11].

Finally, for each spectral broadening regime, the results are presented in Table 1. Additionally, it should be noted that, while the OPA delivered a clean beam profile at a wavelength of 2000 nm, the output from the Herriott cell appeared distorted [Fig. 6]. However, when the beam profile was recorded after the Herriott cell for a wavelength of 2120 nm, the beam exhibited a clean beam profile with only a bit of ellipticity. As the mode-matching and alignment of the Herriott cell could be only performed at a wavelength of 2120 nm, we attribute the beam profile distortion to a change in the OPA mode and a misalignment of the mode-matching. Furthermore, we observed a significant deterioration of the linear polarization in the beam after the Herriott cell for an OPA wavelength of 2000 nm.

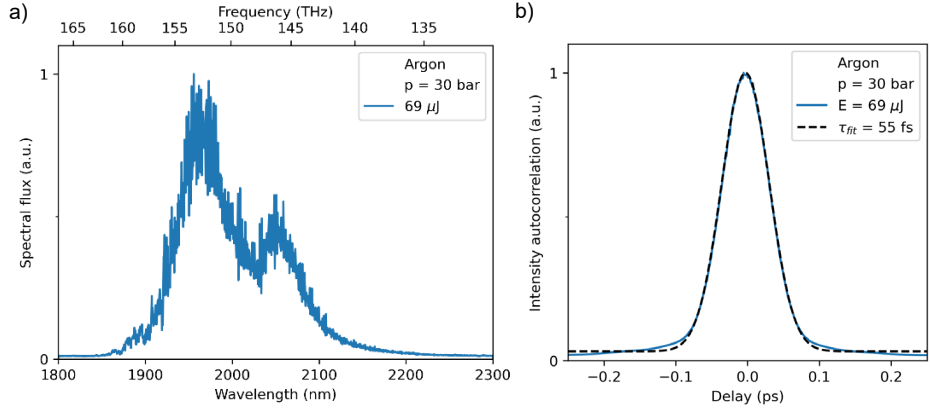

Fig. 11. a) Measured spectrum of the output beam of the Herriott cell for an OPA center wavelength of 2000 nm, and a pulse energy of 69 µJ before the Herriott cell. The Herriott cell contained an argon pressure of 30 bar. The Fourier-transform limited (FTL) pulse duration supported by this spectrum corresponds to about 43 fs. Note that this spectrum was recorded for a different input polarization than the 69 µJ spectrum of Fig. 10. The broadening was slightly reduced. b) The pulse duration was evaluated with an intensity autocorrelator after compressing the pulse with eight bounces on dispersive mirrors (-300 fs$^2$) and 30 mm of sapphire. A Gaussian fit of the autocorrelation assumed a pulse duration of 55 fs (FWHM).

Table 1. Summarized results for pulse compression at the 2 µm spectral range

| Wavelength (nm) | 2120 | 2000 | |
|---|---|---|---|
| Input pulse energy (µJ) | 50 | 58 | 69 |
| Transmission of the Herriott cell (%) | | 80 - 85 | |
| Argon pressure (bar) | 8 | 10 | 30 |
| Broadening regime | Self-compression | Positive dispersion | |
| FTL$^a$ pulse duration of the spectrum (fs) | 40 | 56 | 43 |
| Output pulse duration (fs) | < 46 | 236 (uncompressed) | 55 (compressed) |

$^a$ Fourier-transform limited.

## 4. Discussion and summary

To address the discussed challenges for spectral broadening in gas-filled Herriott cells at wavelengths around 2 µm, we presented an analytical approach to optimize the Herriott cell configuration for strong spectral broadening at a given peak power. Furthermore, this approach allows the use of more compact gas vessels with higher pressures. Overall, this approach could theoretically lead to about three times higher nonlinear phase shifts than in the often-used longest configuration ($M = N - 1$), close to the concentric resonator. This finding also applies to the development of gas-filled Herriott cells for pulse compression of low-energy pulses in the prominent 1 µm wavelengths range.

Experimentally, we demonstrated the temporal compression of 2 µm laser pulses from a commercial optical parametric amplifier (OPA) in two different dispersion regimes. These regimes were accessed by tuning the wavelength of the OPA to two different wavelengths, where the Herriott cell mirrors exhibited second-order dispersion of opposing signs. At a wavelength of 2120 nm and an argon pressure of 8 bar, laser pulses with an energy of about 50 µJ, corresponding to a peak power of about 240 MW, experienced self-compression to a pulse duration below 46 fs. However, strong nonlinear effects led to a beginning breakup of the pulse, which became visible in the pronounced pedestal of the intensity autocorrelation. Spectral broadening in the positive dispersion regime led to an almost ideal Gaussian autocorrelation. The pulses with an energy of 69 µJ could be compressed to a duration of 55 fs. In the future, new mirrors with a flat second order dispersion and direct compensation of the initial chirp of the laser source could potentially improve the results further. The adoption of an ROC of 100 mm, instead of 300 mm, for the Herriott cell mirrors, as discussed previously, will theoretically reduce the required peak power for efficient temporal compression by a factor of almost two. One trade-off of the presented approach comes from the losses accumulated over the large number of reflections. Mirrors with higher reflectivity would be required to increase cell transmission above 85 % for the current number of passes.

The presented considerations and results of pulse compression in the 2 µm range could help to achieve pulse compression of various commercial optical parametric amplifiers and high repetition rate holmium-based regenerative amplifiers with gas-filled Herriott cell. Even Cr:ZnS-based regenerative amplifiers with higher repetition rates might benefit from a recompression of the output pulses, if the bandwidth is not directly reaching the full gain bandwidth.

**Funding.**

The project was funded by Helmut Schmidt University and by n2-Photonics. A.T. acknowledges financial support from the European Research Council under the ERC SoftMeter no. 101076500. Views and opinions expressed are however those of the author(s) only and do not necessarily reflect those of the European Union or the European Research Council Executive Agency.

**Acknowledgment.**

We acknowledge support from Andrei Naumov, who initiated this project by indicating the demand for pulse compression of the output from commercially available 2 µm OPA sources.

**Disclosures.**

Christian Franke and Oleg Pronin declare a conflict of interest. The work was partially financed by n2-Photonics GmbH, which commercializes the pulse-shortening technology. The company was co-founded by Oleg Pronin and Christian Franke.

**Data availability.**

Data underlying the results presented in this paper are not publicly available at this time but may be obtained from the authors upon reasonable request.